\documentclass[12pt]{article}
\usepackage{graphicx,graphics}% Include figure files
\usepackage{amssymb}
\usepackage{amsmath}
\usepackage{dcolumn}% Align table columns on decimal point
\usepackage{bm}% bold math
\usepackage{color}
\usepackage{cite}
\def\be{\begin{equation}}
\def\ee{\end{equation}}
\newcommand{\bea}{\begin{eqnarray}}
\newcommand{\eea}{\end{eqnarray}}
\newcommand{\nn}{\nonumber}

\def\hbar#1{\backslash\hspace{-2mm}#1}

\def\nn{\nonumber}

\def\2tvec#1#2{
\left(
\begin{array}{c}
#1  \\
#2  \\
\end{array}
\right)}

\def\mat2#1#2#3#4{
\left(
\begin{array}{cc}
#1 & #2 \\
#3 & #4 \\
\end{array}
\right) }

\def\Mat3#1#2#3#4#5#6#7#8#9{
\left(
\begin{array}{ccc}
#1 & #2 & #3 \\
#4 & #5 & #6 \\
#7 & #8 & #9 \\
\end{array}
\right) }

\def\3tvec#1#2#3{
\left(
\begin{array}{c}
#1  \\
#2  \\
#3  \\
\end{array}
\right)}

\def\hbar#1{\backslash\hspace{-2mm}#1}

\def\eqn#1{
\begin{eqnarray}
#1
\end{eqnarray}
}
\def\nn{\nonumber}

\newcommand{\bt}{\begin{itemize}}
\newcommand{\et}{\end{itemize}}
\numberwithin{equation}{section}

\begin{document}

\begin{titlepage}
\begin{flushright}
KANAZAWA-12-05, P12039
%March 2012
\end{flushright}

\begin{center}

\vspace{1cm}
{\large\bf Fermionic Dark Matter in Radiative Inverse Seesaw Model with $U(1)_{B-L}$}
\vspace{1cm}

Hiroshi Okada,$^{a}$\footnote{HOkada@kias.re.kr}
Takashi Toma$^{b}$\footnote{t-toma@hep.s.kanazawa-u.ac.jp}
\vspace{5mm}

{\it%
$^{a}${School of Physics, KIAS, Seoul 130-722, Korea}\\
$^{b}${ Institute for Theoretical Physics, Kanazawa University, Kanazawa, 920-1192, Japan}
}
  
  \vspace{8mm}

\abstract{
 We construct a radiative inverse seesaw model with local $B-L$
 symmetry, and investigate the flavor structure of the lepton sector and
 the fermionic Dark Matter. Neutrino masses are radiatively generated through a kind of
 inverse seesaw framework. 
 The PMNS matrix is derived from each mixing matrix of the neutrino and
 charged lepton sector with large Dirac CP phase. 
 We show that the annihilation processes via the interactions with
 Higgses which are independent on the lepton flavor violation, have to
 be dominant in order to satisfy the observed relic abundance by
 WMAP. The new interactions with Higgses allow us to be consistent with
 the direct detection result reported by XENON100, and it is possible to
 verify the model by the exposure of XENON100 (2012). 
 }

\end{center}
\end{titlepage}

\setcounter{footnote}{0}

\section{Introduction}
Inverse seesaw mechanism which generates neutrino masses due to small
lepton number violation is one of the intriguing way to describe tiny
neutrino masses~\cite{Mohapatra:1986aw, Mohapatra:1986bd}. 
Thus interesting phenomenological implications
have been accommodated \cite{non-susy-inverse-pheno1,
non-susy-inverse-pheno2, Khalil:2011tb}.
However Dark Matter (DM) candidate has to be
introduced independently if we discuss DM phenomenology in this kind of
models. 
On the other hand, radiative seesaw mechanism is possible to relate
neutrino mass generation with the existence of DM~\cite{Ma:2006km,
Krauss:2002px, Aoki:2008av}. In particular, the radiative seesaw model
which is proposed by Ma~\cite{Ma:2006km} 
is the simplest model with DM candidates. Subsequently there are a lot of recent works of
the model~\cite{Schmidt:2012yg, Bouchand:2012dx, Ma:2012ez} and the extended
models~\cite{Aoki:2011he, Ahn:2012cg, Farzan:2012sa, Bonnet:2012kz, Kumericki:2012bf,
Kumericki:2012bh, Ma:2012if, Gil:2012ya}. The other radiative neutrino mass models
are studied in Refs.~\cite{Aoki:2010ib, Kanemura:2011vm, Lindner:2011it,
Kanemura:2011mw}. The $\mathbb{Z}_2$ parity 
imposed to the model forbids to have the Dirac neutrino masses, 
produces neutrino masses at one loop level, and stabilizes DM candidates. 
Therefore the feature of the radiative seesaw model motivates to connect
the existence of DM with the neutrino mass generation due 
to inverse seesaw mechanism. 

In this paper, we construct a radiative inverse seesaw model with
$U(1)_{B-L}$ as a concrete example and analyze the neutrino masses,
mixing and the feature of DM. We add three pairs of $B-L$ charged
fermions and a scalar to Ma model~\cite{Ma:2006km}. The scalar particle
breaks the $B-L$ symmetry spontaneously. In the model, the small
Majorana mass terms which violate $U(1)_{B-L}$ weakly and explain the
tiny neutrino masses are generated from a higher operator when the
$U(1)_{B-L}$ symmetry is spontaneously broken. 
We assume the
structure of neutrino mass matrix that induces the best fit value of $\theta_{12}$
of the Pontecorvo-Maki-Nakagawa-Sakata (PMNS) neutrino mixing matrix \cite{mns}:
$\sin^2\theta_{12}$=0.311, which is within $1\sigma$ confidence
level in the global analysis \cite{Tortola:2012te}. Moreover, we introduce
the charged-lepton mixing ($\lambda$) with nonzero Dirac CP phase to
induce non-zero 
$\theta_{13}$ recently reported by T2K \cite{t2k}, Double
Chooz~\cite{Abe:2011fz}, Daya-Bay \cite{daya},
and RENO \cite{reno}. This method was firstly proposed by S. King
\cite{king, King:2012vj}. Due to the mixing $\lambda$, neutrino Dirac Yukawa couplings 
are strongly constrained by the lepton flavor violation (LFV), especially,
$\mu\to e\gamma$ process.
Hence in the original radiative seesaw model~\cite{Ma:2006km}, it is
hard to derive the observed relic density of DM \cite{wmap} 
associated to the annihilation channel if we assume that the lightest
right-handed neutrino is DM. However since the radiative
inverse seesaw model has the other channels with different coupling due to the
additional Higgs boson coming from the $B-L$ symmetry, the observed relic
abundance can be naturally obtained via the channel. Since the additional Higgs boson
mixes with the SM like Higgs, the direct detection with Higgs portal
also comes into the target of a discussion whether the result can be
consistent with the experimental limit reported by XENON100
\cite{xenon}. It implies that the model can be investigated in the direct
search of DM near future. Therefore it is favored to be Higgs portal DM
in the radiative inverse seesaw model from the consistency with
neutrino masses, mixing, LFV and the DM property. 

This paper is organized as follows. In Section 2, we construct the
radiative inverse seesaw model and its Higgs sector. 
In Section 3, we discuss the constraints from LFV, especially, $\mu\to
e\gamma$ process and the DM relic abundance.
In Section 4, we analyze the direct detection of our DM including all the other constraint.
We summarize and conclude the paper in Section 5.

%%%%%%%%%%%%%%%%%%%%%%%%%%%%%%%%%%%%%%
\section{The Radiative Inverse Seesaw Model}
\subsection{Neutrino Mass and Mixing}

\begin{table}[thbp]
\centering {\fontsize{10}{12}
\begin{tabular}{||c|c|c|c|c|c|c|c||c|c|c||}
\hline\hline ~~Particle~~ & ~~$Q$~~ & ~~$u^c, d^c $~~ & ~~$L$~~ & ~~$e^c$~~ & ~~ $N^c$~~
 & ~~$S$~~ & ~~$S'$~~ & ~~$\Phi~~ $& ~~$\eta~~ $ & $\chi $\\\hline
$SU(2)_L$&$\bm{2}$ & $\bm{1}$ & $\bm{2}$ & $\bm{1}$ & $\bm{1}$ &
$\bm{1}$ & $\bm{1}$ & $\bm{2}$ & $\bm{2}$ & $\bm{1}$\\\hline
$Y_{B-L}$ & 1/3 & -1/3 & -1 & 1 & 1 & -1/2 & 1/2 & 0 & 0 & -1/2\\\hline
%Spin&1/2&1/2&1/2&1/2&1/2&1/2&1/2&0&0&0\\\hline
$\mathbb{Z}_2$ & + & + & + & + & $-$ & $-$ & + & + & $-$ & +\\
\hline\hline
\end{tabular}%
} \caption{The field contents and the charges. Three generations of right handed neutrinos
 $N_i^c$ and additional fermions $S_i$, $S'_i$ are introduced. A 
pair of fermions $S$ and $S'$ is required to cancel the
anomaly. The SM, inert and $B-L$ Higgs bosons are denoted by $\Phi$, $\eta$, and
$\chi$, respectively. } \label{tab:b-l}
\end{table}

The radiative inverse seesaw mechanism can be realized by introducing
the $B-L$ symmetry which is spontaneously broken at TeV
scale~\cite{non-susy-inverse-pheno2}. The field contents of our model is
shown in Table~1 and the Lagrangian in the lepton sector is
\begin{equation}
{\cal L}=
y_\ell  L\Phi e^c +y_\nu  L\eta N^c+y_S N^c\chi S +{\lambda_{S}
\over \Lambda} \chi^{\dagger2} S^2 + {\lambda_{S'} \over
\Lambda} \chi^{2} {S'}^2 + \mathrm{h.c.}, 
\label{eq:lagrangian}
\end{equation}
where $\Lambda$ is a cut-off scale and the generation indices are
abbreviated. Note that the mass term
$SS'$ can be forbidden by the $Z_2$ symmetry\footnote{Another way to
induce the mass term of $S$ is proposed by E. Ma \cite{Ma:2009gu}, in
which the term is done at loop level.}. 
%%%
 We have another dimension 5 operator such as $L\Phi
S'\chi^\dag$, which affects on the neutrino mass, and it would be
difficult to forbid it.  Thus we need to assume the coupling of this operator
is small enough.
%The mass scale after the breaking is of course 100 times as small as the scale of $\mu$.  

%%%
After the symmetry breaking, that is $\chi=(\chi^0+v')/\sqrt2$
%%%b
\footnote{
The value of $v'$ can be constrained by the LEP experiment that tells us $m_{Z'}/g' =|Y^\chi_{B-L}| v' > 6$ TeV \cite{Carena:2004xs}, where $m_{Z'}$ and $g'$ are the $B-L$ gauge boson and the $B-L$ guage coupling, respectively. Since Y$^\chi_{B-L}$=-1/2 is taken in our case, $v' >$ 12 TeV is obtained.},
%%%
$\phi=(\phi^0+v)/\sqrt2$ with $\Phi=(\phi^+,\phi)^T$ and $v=246\
{\rm GeV}$, the neutrino sector in the flavor basis can be written as
\begin{equation}
{\cal L}_m^{\nu} = y_\nu  L\eta N^c + M^TN^c S+ \frac{\mu}{2}S^2
+\mathrm{h.c.}, 
\end{equation}
where $ M =y_{S} v'/\sqrt{2}$ and $\mu = \lambda_{S}v'^2/\Lambda$. 
The scale of $\mu\sim1~\mathrm{keV}$ which corresponds to
$\Lambda\sim10^{14}~\mathrm{GeV}$ is required as we will see in the following section.
The inert doublet $\eta\equiv(\eta^+,(\eta_R+i\eta_I)/\sqrt2)^T$ does not
have any vacuum expectation values (VEV). As a result, the $6\times 6$ neutrino mass
matrix in the basis $(N^c, S)$ takes the form:%
\be
{\cal M}_\nu
=
\left(
\begin{array}{cc}
   0 & M^T \\
  M & \mu\\
\end{array}
\right).
\ee
 The mass matrix $M$ and $\mu$ cannot be diagonalized
simultaneously in general. We assume that the mass matrix $M$ and $\mu$ are diagonal
as $M=\mathrm{diag}\:(M_1,M_2,M_3)$ and
$\mu=\mathrm{diag}\:(\mu_1,\mu_2,\mu_3)$ for simplicity.
Then the $6\times6$ mass matrix $\mathcal{M}_{\nu}$ can be diagonalized
for every
family as $\mathrm{diag}(m_{i+},m_{i-})$ by the unitary matrix $U_i$ for $i$-th
family. 
The mass eigenvalues are expressed as
\begin{equation}
m_{i\pm}=\sqrt{M_i^2+\frac{\mu_i^2}{4}}\pm\frac{\mu_i}{2}.
\end{equation}
When $\mu_i\ll M_i$, the neutrino masses are degenerated. 
The unitary matrix $U_i$ can be expressed as
\begin{eqnarray}
U_i=
 \left(
\begin{array}{cc}
\frac{M_i}{\sqrt{M_i^2+m_{i+}^2}}&
\frac{iM_i}{\sqrt{M_i^2+m_{i-}^2}}\\
\frac{m_{i+}}{\sqrt{M_i^2+m_{i+}^2}}&
-\frac{im_{i-}}{\sqrt{M_i^2+m_{i-}^2}}\\
\end{array}
\right),
\end{eqnarray}
where $(m_{i+},m_{i-})=U_i^T{\cal M}_\nu^{i} U_i$ and
$\mathcal{M}_{\nu}^i$ implies the $2\times2$ mass matrix for $i$-th family.
The flavor eigenstates are rewritten by the mass eigenstates $\nu_{i\pm}$ as follows:
\begin{eqnarray}
N^c_i&=&\frac{M_i}{\sqrt{M_i^2+m_{i+}^2}}\nu_{i+}
+\frac{iM_i}{\sqrt{M_i^2+m_{i-}^2}}\nu_{i-},\\
S_i&=&\frac{m_{i+}}{\sqrt{M_i^2+m_{i+}^2}}\nu_{i+}
-\frac{im_{i-}}{\sqrt{M_i^2+m_{i-}^2}}\nu_{i-}.
\end{eqnarray}
The light neutrino mass matrix seen in Fig.~\ref{fig:mass} is given 
as Ref.~\cite{Ma:2006km} by 1-loop radiative correction: 
\begin{figure}[t]
\begin{center}
\includegraphics[scale=1]{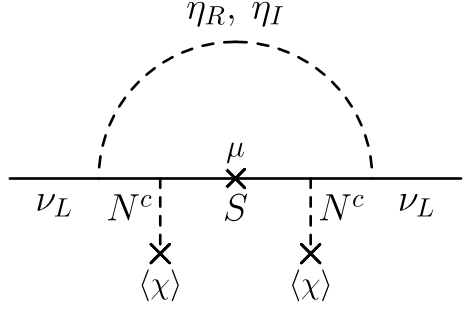}
\caption{Neutrino mass generation via radiative inverse seesaw.}
\label{fig:mass}
\end{center}
\end{figure}
\begin{eqnarray}
\left(m_\nu\right)_{\alpha\beta}
&=&-\sum_{i=1}^3\frac{\left(y_{\nu}\right)_{\alpha i}\left(y_{\nu}\right)_{\beta i}}{(4\pi)^2}
\frac{M_{i}^2m_{i-}}{M_i^2+m_{i-}^2}
\left[\frac{m^2_R}{m^2_R-m_{i-}^2}\log\frac{m^2_R}{m_{i-}^2}
-\frac{m^2_I}{m^2_I-m_{i-}^2}\log\frac{m^2_I}{m_{i-}^2}\right]\nonumber\\
&&+\sum_{i=1}^3
\frac{\left(y_{\nu}\right)_{\alpha i}\left(y_{\nu}\right)_{\beta
i}}{(4\pi)^2}
\frac{M_{i}^2m_{i+}}{M^2+m_{i+}^2}
\left[\frac{m^2_R}{m^2_R-m_{i+}^2}\log\frac{m^2_R}{m_{i+}^2}
-\frac{m^2_I}{m^2_I-m_{i+}^2}\log\frac{m^2_I}{m_{i+}^2}\right],
\end{eqnarray}
where $m_R$ and $m_I$ imply masses of $\eta_{R}$ and $\eta_{I}$.
When $\mu_i\ll M_i$, we can obtain the approximate light neutrino mass
matrix by expanding $m_{i\pm}$ up to the leading order, 
\begin{eqnarray}
\left(m_{\nu}\right)_{\alpha\beta}
\!\!\!&\simeq&\!\!\!
-\sum_{i=1}^3
\frac{\left(y_{\nu}\right)_{\alpha i}\left(y_{\nu}\right)_{\beta i}\mu_i}{2(4\pi)^2}\left[
\frac{m_R^2}{M_i^2}I\left(\frac{M_i^2}{m_R^2}\right)
-\frac{m_I^2}{M_i^2}I\left(\frac{M_i^2}{m_I^2}\right)
\right]+\mathcal{O}\left(\mu_i^2\right),
\label{eq:neut-mass0}
\end{eqnarray}
where we will define the function $I(x)$ and the parameter $\Lambda_i$ as
\begin{equation}
I(x)=\frac{x}{1-x}\left(1+\frac{x\log{x}}{1-x}\right),\ 
\Lambda_i=
\frac{\mu_i}{2(4\pi)^2}\left[
\frac{m_R^2}{M_i^2}I\left(\frac{M_i^2}{m_R^2}\right)
-\frac{m_I^2}{M_i^2}I\left(\frac{M_i^2}{m_I^2}\right)
\right].
\label{eq:loop-int}
\end{equation}

%%%
We can see from Eq.~(\ref{eq:neut-mass0}) that the mixing
matrix of the light neutrino mass matrix is determined by the structure
of the neutrino Yukawa matrix $y_{\nu}$ since the majorana mass matrix
$\mu$ is assumed to be diagonal in this case. 
In order to identify the structure of $y_{\nu}$, here we set a specific
texture of the neutrino mass matrix, which induces
the best fit values of $\theta_{12}$, as
%The above texture can lead to a following neutrino mass matrix: 
\begin{equation}
m_{\nu}=\left(
\begin{array}{ccc}
A & B & -B\\
B & (3A+\sqrt3B)/6 & -(3A+\sqrt3B)/6\\
-B & -(3A+\sqrt3B)/6 & (3A+\sqrt3B)/6
\end{array}
\right).
\label{eq:neut-mass}
\end{equation}
Then the mass matrix can be diagonalized by the following mixing matrix and
the eigenvalues are written as 
\eqn{O_{\nu L}=
\Mat3{\sqrt{(1+1/\sqrt7)/2} } {-\sqrt{(1-1/\sqrt7)/2}} {0} 
{-\sqrt{1-1/\sqrt7}/2} {-\sqrt{1+1/\sqrt7}/2} {1/\sqrt{2}}
{\sqrt{1-1/\sqrt7}/2} {\sqrt{1+1/\sqrt7}/2} {1/\sqrt{2}}
\label{mns},}
\eqn{
m_1=A+\frac{2B}{\sqrt{3}}\left(1-\sqrt{7}\right),\
m_2=A+\frac{B}{\sqrt{3}}\left(1+\sqrt{7}\right),\
m_3=0.
\label{eq:mass-eigenvalue}
}
Thus the squared mass differences are $\Delta m^2_{{\bf sol}}\equiv
m^2_2-m^2_1$ and  $\Delta m^2_{{\bf atm}}\equiv |m^2_1-m^2_3|$, 
and the neutrino mass hierarchy is predicted to be inverted.
% is given by
%\eqn{
%\Delta m^2_{{\bf sol}}=4B\sqrt{\frac{7}{3}}\left(A+\frac{B}{\sqrt{3}}\right),\ \Delta m^2_{{\bf
%atm}}=m^2_1=(A-2\sqrt{3}B/(1+\sqrt7))^2.
%\label{eq:mass-eigenvalue2}
%}
In order to get $\Delta
m^2_{{\bf sol}}=7.62\times10^{-5}$ eV$^2$, $\Delta m^2_{{\bf
atm}}=2.40\times10^{-3}$ eV$^2$, which are the best fit values
\cite{Tortola:2012te}, we find the following solutions:
\eqn{
(A,B) &=& (\pm4.92\times10^{-2},\:\:\pm2.53\times10^{-4}),\quad 
(\pm1.83\times10^{-2},\:\:\mp3.23\times10^{-2})~\mathrm{eV}.
\label{eq:AB}
} 
Notice that the other solutions do not exist any more. 
We cannot obtain non-zero $\sin\theta_{13}$ from the light neutrino mass
matrix Eq.~(\ref{eq:neut-mass}). Non-zero $\sin\theta_{13}$ is derived
from the charged lepton mixing as we will discuss below. 
We find $\sum_{i} m_{i}\simeq7.7\times10^{-4}$ eV, and
the effective mass $\left<m_{ee}\right>\equiv |\sum_{i}{\left(O_{\nu L}\right)}_{1i}^2m_{i}|\simeq0.026$
eV and $\tan\theta_{12}=\left(1-\sqrt{7}\right)/\sqrt{6}$
which gives $\sin^2\theta_{12}=0.311$ which is the best fit 
$1\sigma$ value~\cite{Tortola:2012te}. The recent experimental value
for $\left<m_{ee}\right>$ are referred in Ref.~\cite{Rodejohann:2011mu}. 
We can choose the following texture which leads the above neutrino mass 
matrix\footnote{Our parametrization is taken so that DM $\nu_{1-}$ couples to
the charged leptons. Even if one selects
another parametrization, {\it e.g}, by replacing the first column
and the second one of $y_\nu$, the severe constraints from LFV do not change
as we will discuss in the next section. }
: 
\eqn{
y_\nu=
\Mat3{0}{0}{b} {a}{0}{c} {-a}{0}{-c}
,
\label{neu-y}}
where the parameters $a$, $b$, $c$ are expressed by $A$ and $B$ as follows:
\eqn{
a&=&\frac{1}{\sqrt{2A\Lambda_1}}
\sqrt{\left(A+\frac{1+\sqrt{7}}{\sqrt{3}}B\right)\left(A+\frac{1-\sqrt{7}}{\sqrt{3}}B\right)},
\ b=\frac{A}{\sqrt{A\Lambda_3}},\ 
c=\frac{B}{\sqrt{A\Lambda_3}
\label{eq:parameters}
}.}
 
\begin{figure}[tbc]
\begin{center}
\includegraphics[scale=0.8]{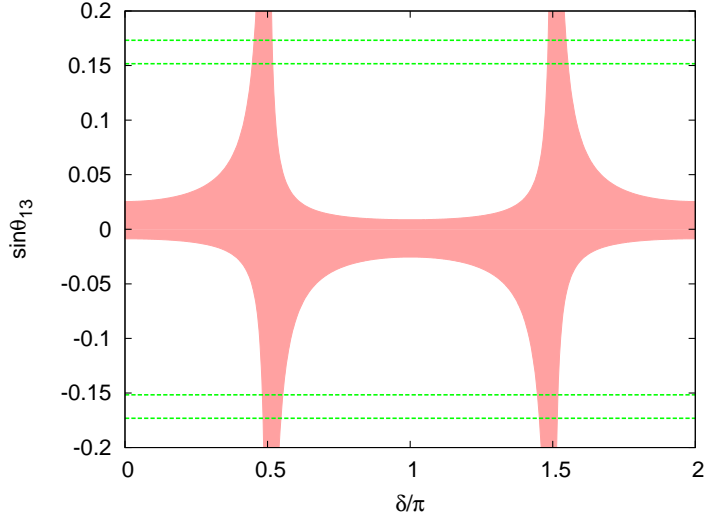}
\caption{$\sin\theta_{13}$ versus CP Dirac phase $\delta$. Here we
 restrict $\sin\theta_{12}$ within the range of
 the best fit with 1$\sigma$ that is $0.303\le\sin^2\theta_{12}\le0.335$
 \cite{Tortola:2012te}.}
\label{13-delta}
\end{center}
\end{figure}

To induce non-zero $\theta_{13}$, we consider the charged lepton mixing
\cite{king, King:2012vj}.
If we set the Dirac mass matrix of charged leptons $m_{e}$ and the mixing matrix
$U_{eL}$ as
\eqn{
m_e=\frac{v}{\sqrt2}
\Mat3{y^{\ell}_1} {y^{\ell}_2} {0} 
{y^{\ell}_2} {y^{\ell}_3} {0}
{0}{0}  {y_4^{\ell}}
\label{chgd},\quad
U_{eL}\sim
\Mat3{1} {\lambda e^{i\delta}} {0} 
{-\lambda e^{-i\delta}} {1} {0}
{0}{0}  {1}
\label{chgd-mix},
}
where we define
$(|m_e|^2,|m_\mu|^2,|m_\tau|^2)=U^\dag_{eL}m_em^\dag_eU_{eL}$ and
$\delta$ is the Dirac CP phase.
From the mixing matrix $O_{\nu L}$ and $U_{eL}$, we can obtain 
each of the element of the PMNS matrix, which is defined as $U_{PMNS}= U^\dag_{eL}O_{\nu}P$,
is found by
\eqn{
\sin\theta_{13}\simeq-\frac{\lambda}{\sqrt2}, \ \sin\theta_{12}\simeq
-\sqrt{(1-1/\sqrt7)/2}+\frac{\lambda}{\sqrt2}\sqrt{(1+1/\sqrt7)/2}\cos\delta, \ 
\sin\theta_{23}\simeq \frac{1}{\sqrt2},
\label{eq:sin13}
}
where $P$ contains two majorana phases. 
The allowed value of $\sin\theta_{13}$ is shown in
Fig.~\ref{13-delta} as the function of $\delta$, within the range of the best fit
with 1$\sigma$ that is
$0.303\le\sin^2\theta_{12}\le0.335$~\cite{Tortola:2012te}. 
The light red region is in 1$\sigma$ range of $\sin\theta_{12}$. 
It suggests that
large CP Dirac phase is required to satisfy the current
global experimental limit of $\sin\theta_{13}$ for inverted hierarchy 
such as $0.023\leq\sin^2\theta_{13}\leq0.030$~\cite{Tortola:2012te}
which is shown as the two green sandwiched regions.
%Each experimental limit is shown for Double Chooz~\cite{Abe:2011fz},
%Daya-Bay~\cite{daya}, and RENO~\cite{reno}.
%%%%%%%%
%%%%%%%%

\subsection{Higgs Sector}
%The Higgses $\phi^0$ and $\chi^0$ mix after the symmetry breaking. These mass eigenstates are relative with the annihilation and direct detection of DM via the Higgses exchange.
The Higgses $\phi^0$ and $\chi^0$ mix after the symmetry breaking and these are not mass eigenstates. Interactions should be written by mass eigenstates in order to analyze the DM relic density and direct detection in the next section.
%%%
The Higgs potential of this model is given by
\begin{eqnarray}
V \!\!\!&=&\!\!\!
 m_1^{2} \Phi^\dagger \Phi + m_2^{2} \eta^\dagger \eta  + m_3^{2} \chi^\dagger \chi +
\lambda_1 (\Phi^\dagger \Phi)^{2} + \lambda_2 
(\eta^\dagger \eta)^{2} + \lambda_3 (\Phi^\dagger \Phi)(\eta^\dagger \eta) 
+ \lambda_4 (\Phi^\dagger \eta)(\eta^\dagger \Phi)
\nonumber \\ &&\!\!\!
+
\lambda_5 [(\Phi^\dagger \eta)^{2} + \mathrm{h.c.}]+
\lambda_6 (\chi^\dagger \chi)^{2} + \lambda_7  (\chi^\dagger \chi)
(\Phi^\dagger \Phi)  + \lambda_8  (\chi^\dagger \chi) (\eta^\dagger \eta),
\end{eqnarray}
where $\lambda_5$ has been chosen real without any loss of
generality. The couplings $\lambda_1$, $\lambda_2$ and $\lambda_6$ have to be positive
to stabilize the potential.  
Inserting the tadpole conditions; $m^2_1=-\lambda_1v^2-\lambda_7v'^2/2$
and $m^2_3=-\lambda_6v'^2 - \lambda_7v^2/2$, 
the resulting mass matrices are given by
\begin{eqnarray}
m^{2} (\phi^{0},\chi^0) = \left(%
\begin{array}{cc}
  2\lambda_1v^2 & \lambda_7vv' \\
  \lambda_7vv' & 2\lambda_6v'^2 \\
\end{array}%
\right) &=& \left(\begin{array}{cc} \cos\alpha & \sin\alpha \\ -\sin\alpha & \cos\alpha \end{array}\right)
\left(\begin{array}{cc} m^2_{h} & 0 \\ 0 & m^2_{H}  \end{array}\right)
\left(\begin{array}{cc} \cos\alpha & -\sin\alpha \\ \sin\alpha & \cos\alpha \end{array}\right), \nn\\\\ 
m^{2} (\eta^{\pm}) &=& m_2^{2} + \frac12 \lambda_3 v^{2} + \frac12 \lambda_8 v'^{2}, \\ 
m^2_R\equiv m^{2} ( Re \eta^{0}) &=& m_2^{2} + \frac12 \lambda_8 v'^{2}
 + \frac12 (\lambda_3 + \lambda_4 + 2\lambda_5) v^{2}, \\ 
m^2_I\equiv m^{2} ( Im \eta^{0}) &=& m_2^{2} + \frac12 \lambda_8 v'^{2}
 + \frac12 (\lambda_3 + \lambda_4 - 2\lambda_5) v^{2},
\end{eqnarray}
where $h$ implies SM-like Higgs and $H$ is an additional Higgs mass
eigenstate. 
The tadpole condition for $\eta$, which is given by
$\left.\frac{\partial V}{\partial \eta}\right|_{\mathrm{VEV}}=0$, tells us that  
\be
m^2_2>0,\ \lambda_2>0,\ \lambda_3+\lambda_4+2\lambda_5>0,\ \lambda_8>0,
\ee
in order to satisfy the condition $v_\eta=0$ at tree level.
The masses of $\phi^0$ and $\chi^0$ are rewritten in terms of the mass eigenstates of $h$ and $H$
as
\begin{eqnarray}
\phi^0 &=& h\cos\alpha + H\sin\alpha, \nn\\
\chi^0 &=&- h\sin\alpha + H\cos\alpha.
\label{eq:mass_weak}
\end{eqnarray}

%%%%%%%%%%%%%%%%%%%%%%
%%%%%%%%%%%%%%%%%%%%%
\section{The Constraints from Lepton Flavor Violation and DM Relic
 Density}
\subsection{Lepton Flavor Violation}
We investigate Lepton Flavor Violation (LFV) under the flavor structure Eq.~(\ref{neu-y}).
%%%
 We put a reasonable approximation $m_{i\pm}\simeq M_i$ hereafter,
since the scale of $\mu$ that is keV scale is
negligible compared to the scale of $M_i$ that is expected ${\cal
O}(10^{2\sim3})$ GeV.
%%%
The experimental upper bounds of the branching ratios are
$\mathrm{Br}\left(\mu\to e\gamma\right)\leq2.4\times 10^{-12}$~\cite{Adam:2011ch},
$\mathrm{Br}\left(\tau\to\mu\gamma\right)\leq4.4\times10^{-8}$ and
$\mathrm{Br}\left(\tau\to e\gamma\right)\leq3.3\times10^{-8}$~\cite{Nakamura:2010zzi}. 
The branching ratios of the processes
$\ell_{\alpha}\rightarrow\ell_{\beta}\gamma$~$(\ell_\alpha,\ell_\beta=e,\mu,\tau)$
are calculated as
 \begin{eqnarray}
  {\rm Br}(\ell_{\alpha}\rightarrow\ell_{\beta}\gamma)
&\!\!\!=\!\!\!&
\frac{3\alpha_{\mathrm{em}}}{64\pi(G_{F}M^{2}_{\eta})^{2}}
\left|\sum_{i=1}^3\left(U_{eL}^{\dag}y_{\nu}^{\dag}\right)_{\alpha{i}}
\Bigl(y_{\nu}U_{eL}\Bigr)_{i\beta}
F_2\left(\frac{M_i^2}{M_{\eta}^2}\right)\right|^{2}{\rm
Br}(\ell_{\alpha}\rightarrow\ell_{\beta}\overline{\nu_{\beta}}\nu_{\alpha}),
 \label{LFV1}
 \end{eqnarray}
where $\alpha_{\mathrm{em}}=1/137$, $\mathrm{Br}\left(\mu\to
e\overline{\nu_e}\nu_\mu\right)=1.0$, $\mathrm{Br}\left(\tau\to
e\overline{\nu_e}\nu_\tau\right)=0.178$,
$\mathrm{Br}\left(\tau\to\mu\overline{\nu_\mu}\nu_\tau\right)=0.174$, 
$M_\eta$ is $\eta^+$ mass, $G_{F}$ is the Fermi constant and 
the loop function $F_{2}(x)$ is given by
\begin{eqnarray}
F_{2}(x)=\frac{1-6x+3x^{2}+2x^{3}-6x^{2}\ln{x}}{6(1-x)^{4}}.
\end{eqnarray}
The $\mu\to e\gamma$ process gives the most stringent constraint. 
If the mixing matrix of the charged leptons
Eq.~(\ref{chgd-mix}) is diagonal, i.e. $\lambda=0$, 
the $\mu\to e\gamma$ process does not constrain the model since the
branching ratio Eq.~(\ref{LFV1}) can be zero. Although
$\tau\to\mu\gamma$ and $\tau\to e\gamma$ processes remain as LFV
constraint, these are much weaker than $\mu\to e\gamma$.
Instead of that, non-zero $\theta_{13}$ is not derived. 
Thus we can insist that $\mu\to e\gamma$ constraint is closely correlated with
non-zero $\theta_{13}$. In order to be consistent with LFV and obtaining
non-zero $\theta_{13}$, the neutrino Yukawa couplings must be small
enough to escape the LFV constraint. 

%%%%%%%%
\subsection{DM Relic Density}
\begin{figure}[t]
\begin{center}
\includegraphics[scale=0.8]{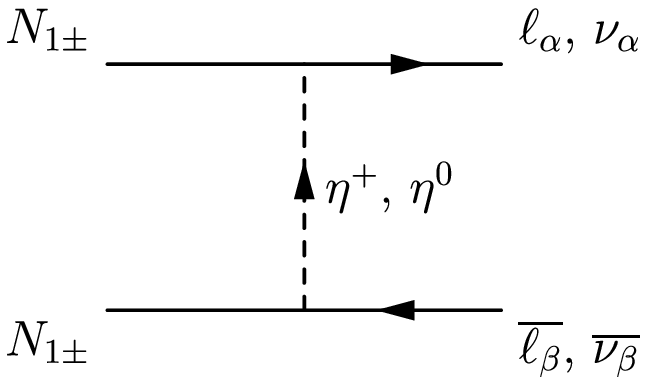}\qquad
\includegraphics[scale=0.8]{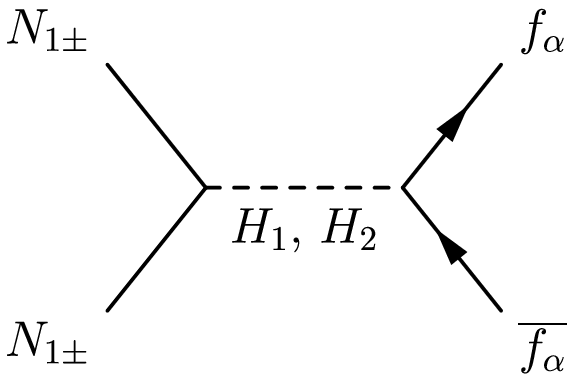}\qquad
\raisebox{0.16cm}{\includegraphics[scale=0.8]{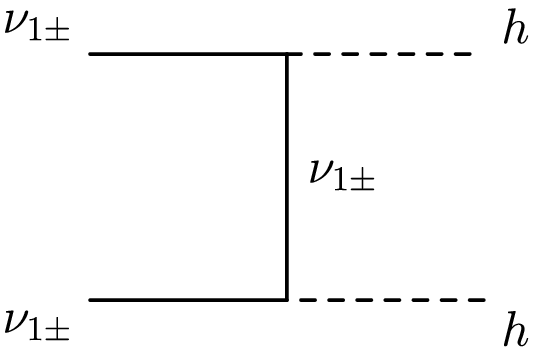}}
\caption{$t$, $u$ and $s$-channel of coannihilation processes of DM $\nu_{1\pm}$.}
\label{fig:ann}
\end{center}
\end{figure}

If we assume that DM is fermionic and mass hierarchy $M_1<M_2<M_3$
for the right-handed neutrinos $N_i$, the mass eigenstates $\nu_{1\pm}$
can be highly degenerated DMs due to the weak lepton number violation
term $\mu_i$ which induces small neutrino masses. Thus we have to take into account
coannihilation of $\nu_{1-}$ and $\nu_{1+}$.
The typical interacting terms are found as
 \begin{eqnarray}
{\cal L}_{\mathrm{int}}
 \!\!\!&\simeq&\!\!\!
\left(U_{eL}y_\nu\right)_{\alpha 1} 
\left(\overline{\ell_{\alpha}}\eta^+ -
 \overline{\nu_{\alpha}}\eta^0\right)\left[-\frac{i}{\sqrt{2}} \nu_{1-}
+\frac{1}{\sqrt{2}}\nu_{1+}\right]+\mathrm{h.c.}\nonumber\\
&&
+\frac{\left(y_S\right)_{11}\sin\alpha}{2}{h}\left(\nu_{1-}^2+\nu_{1+}^2\right)
-\frac{\left(y_S\right)_{11}\cos\alpha}{2}{H}\left(\nu_{1-}^2+\nu_{1+}^2\right)
\nonumber\\
&&+\frac{y_{f}\cos\alpha}{\sqrt{2}}{h}\overline{f}f+\frac{y_f\sin\alpha}{\sqrt{2}}{H}\overline{f}f,
 \end{eqnarray} 
  where the masses of $\eta^{0(\pm)}$ assumed to be always heavier
 than $\nu_{\pm}$ to avoid the too short lifetime of DMs through our analysis. 
Three types of the coannihilation processes exist and these are shown in
 Fig~\ref{fig:ann}. The effective cross section to
 $\ell\overline{\ell}$, $f\overline{f}$ and $hh$ are given as
 \begin{eqnarray}
\sigma^{\ell}_{\mathrm{eff}} v_{\rm rel}
\!\!\!&\simeq&\!\!\!
\frac{1}{96\pi}
\left(\sum_{\alpha=e}^{\tau}\left|\left(U_{eL}y_{\nu}\right)_{\alpha1}\right|^2\right)^2
\frac{M_1^2\left(M_1^4+M_\eta^4\right)}{\left(M_1^2+M_\eta^2\right)^4}v_{\mathrm{rel}}^2,
\label{eq:ann1}
\\
\sigma^{t}_{\mathrm{eff}} v_{\rm rel}
\!\!\!&\simeq&\!\!\!
\frac{3\left(y_S\right)_{11}^2y_t^2 M_1^2}{64\pi} 
\left|\frac{\sin\alpha\cos\alpha}{4M_1^2-m_{h}^2+im_{h}\Gamma_h}
-\frac{\sin\alpha\cos\alpha}{4M_1^2-m_{H}^2+im_{H}\Gamma_H}\right|^2
 \left(1-\frac{m_t^2}{M_1^2}\right)^{3/2}v^2_{\rm rel},
\label{eq:ann2}
\\
\sigma_{\mathrm{eff}}^{h}v_{\mathrm{rel}}
\!\!\!&\simeq&\!\!\!
\frac{\left(y_S\right)_{11}^4\sin^4\alpha{M_1^2}}{32\pi\left(m_h^2-2M_1^2\right)^2}
\left[
1-\frac{1}{3}\frac{m_h^2-M_1^2}{m_h^2-2M_1^2}
+\frac{1}{12}\left(\frac{m_h^2-M_1^2}{m_h^2-2M_1^2}\right)^2
\right]\sqrt{1-\frac{m_h^2}{M_1^2}}\:v_{\mathrm{rel}}^2
 \end{eqnarray}
where $\eta_R$, $\eta_I$ and $\eta^+$ masses are regarded as same
parameter $M_\eta$ for simplicity and only top pair is taken into account in
fermion pair $f\overline{f}$ because of the largeness of the Yukawa coupling. 
%The Higgs $H$ is assumed to be heavier than DMs $\nu_{1\pm}$.
The SM-like Higgs mass and decay width are fixed to
$m_h=125~\mathrm{GeV}$ and $\Gamma_h=10^{-2}~\mathrm{GeV}$, and the
heavy Higgs mass is assumed to be $m_H<200~\mathrm{GeV}$. 
The decay width of the heavy Higgs $\Gamma_H$ is expressed as 
\begin{equation}
\Gamma_H=\frac{y_t^2\sin^2\alpha}{16\pi}m_H\left(1-\frac{4m_t^2}{m_H^2}\right)^{3/2}.
\end{equation}
The contribution of the process $H\to\nu_{1\pm}\nu_{1\pm}$ is also added
to the decay width when the relation $m_H>2M_1$ is satisfied. 
Note that the mixing matrix of charged leptons $U_{eL}$ is introduced in the
effective cross section $\sigma_{\mathrm{eff}}^{\ell}v_{\mathrm{rel}}$ since the initial
Lagrangian~(\ref{eq:lagrangian}) is not
assumed as diagonal base of charged leptons. The s-wave vanishes and
p-wave only remains in the above annihilation cross section due to the
helicity suppression.

Since the neutrino Yukawa couplings $y_{\nu}$ are severely restricted by
LFV, the annihilation cross section of $\ell\overline{\ell}$-channel is
too small to obtain the proper DM relic 
abundance. Thus large contributions from $t\overline{t}$ and $hh$ channels
are required to have sizable effective annihilation cross section. 
These processes are possible because of the mixing of Higgses $\phi^0$
and $\chi^0$, namely the DMs $\nu_{1\pm}$ can be Higgs portal DMs.
This is a different aspect from the loop induced neutrino mass
model~\cite{Ma:2006km} and the various analysis of the right-handed neutrino DM
in the original model~\cite{Schmidt:2012yg, Kubo:2006yx, Kubo:2006rm, Suematsu:2009ww,
Suematsu:2010nd}. 

\begin{figure}[t]
\begin{center}
\includegraphics[scale=0.7]{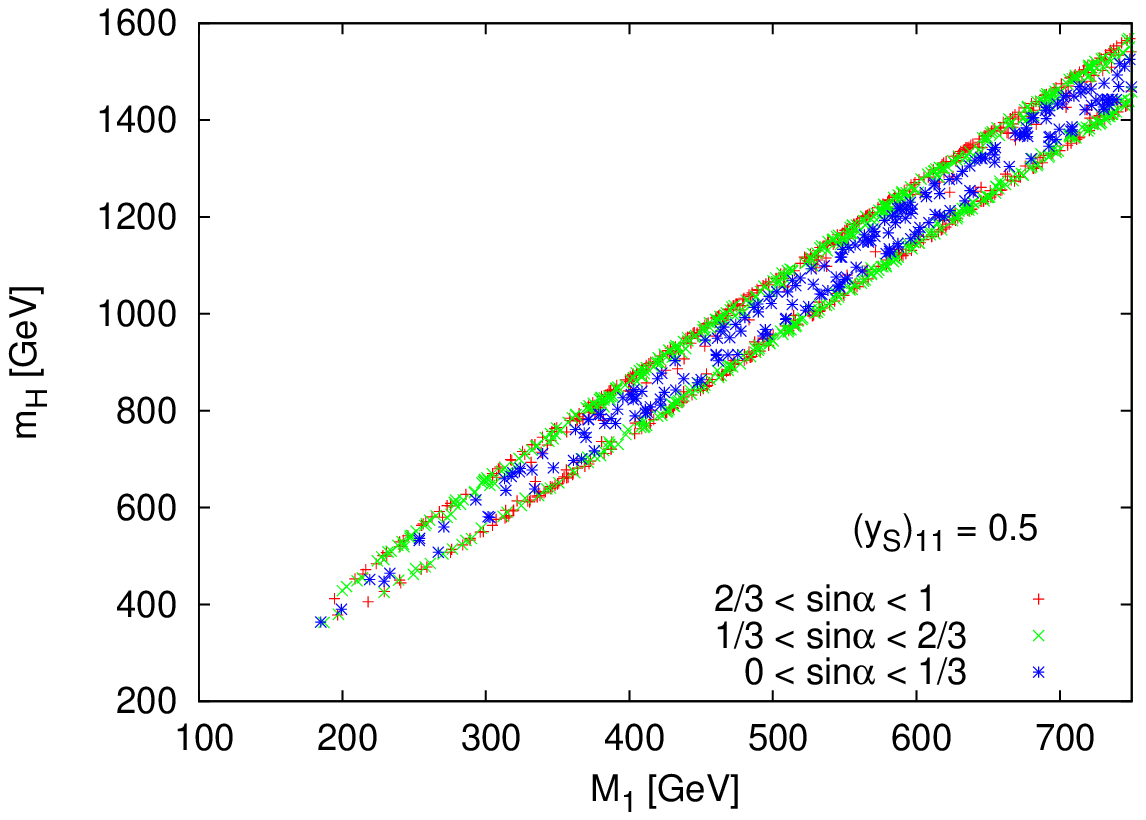}
\includegraphics[scale=0.7]{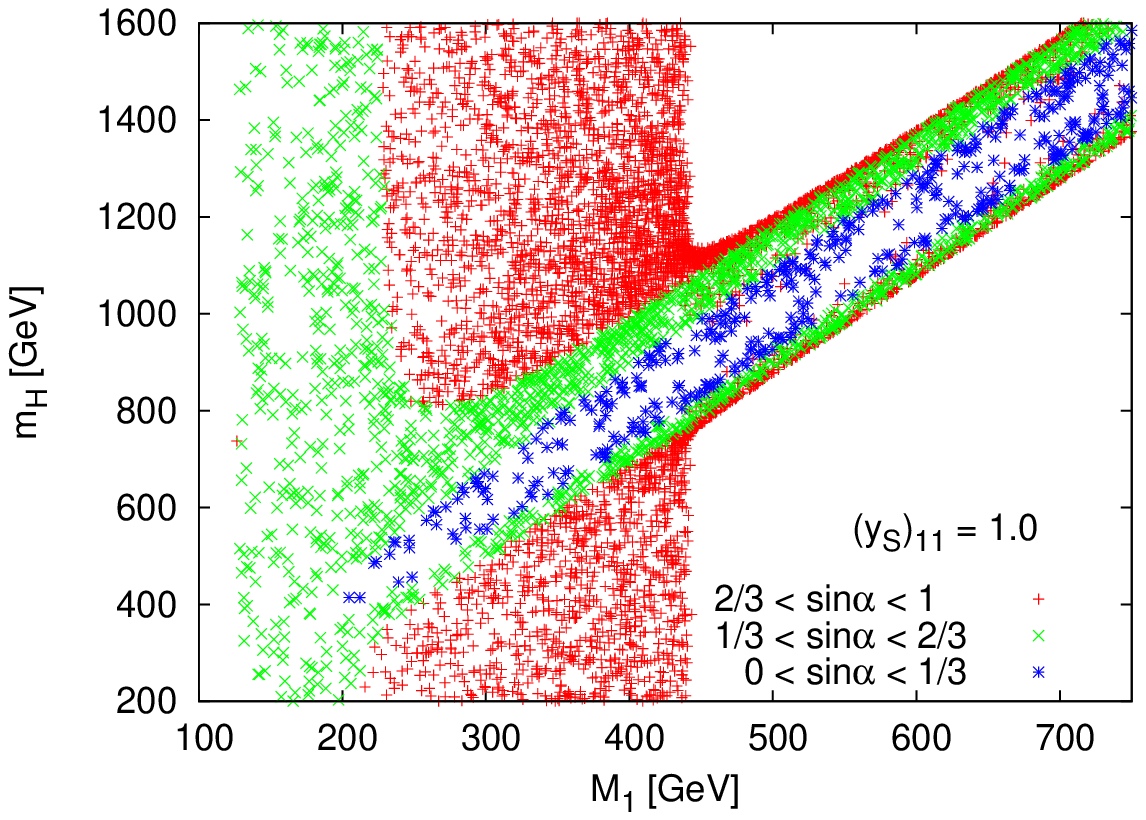}
\caption{The allowed parameter spaces for
 $\left(y_S\right)_{11}=0.5$ and $1.0$ in $M_1$-$m_H$ plane. 
 The other parameters are fixed to $\Lambda_i=1.0~\mathrm{eV}$,
 $M_2=1.5~\mathrm{TeV}$, $M_3=2.0~\mathrm{TeV}$. The unfixed parameters
 are $M_1$, $M_\eta$, $m_H$ and $\sin\alpha$. 
}
\label{fig:plot}
\end{center}
\end{figure}

The independent parameters which appear in the analysis are $\Lambda_1$,
$\Lambda_3$, $M_1$, $M_3$, $M_\eta$, $m_H$, $\left(y_S\right)_{11}$ and
$\sin\alpha$. The parameters $A$ and $B$
of the neutrino mass matrix Eq.~(\ref{eq:neut-mass}) 
are determined by neutrino mass eigenvalues as Eq.~(\ref{eq:AB}), and
we take the fourth solution of Eq.~(\ref{eq:AB}) as example. 
The parameter $\lambda$ of the charged lepton mixing matrix is fixed by the
experimental value of $\sin\theta_{13}$ as Eq.~(\ref{eq:sin13}), and 
$\Lambda_2$ and $M_2$ are not relative with the analysis since the
second column of the neutrino Yukawa matrix Eq.~(\ref{neu-y}) is zero. 
The allowed parameter spaces from LFV and the DM relic abundance for
$\left(y_S\right)_{11}=0.5$ and $1.0$ in $M_1$-$m_H$ plane are shown in
Fig.~\ref{fig:plot} where the parameters are fixed to
$\Lambda_i=1~\mathrm{eV}$, $M_2=1.5~\mathrm{TeV}$ and
$M_3=2.0~\mathrm{TeV}$. 
The parameter choice $\Lambda_i\sim1~\mathrm{eV}$ means that $\mu_i\sim
1~\mathrm{keV}$ if $I\left(M_i^2/m_R^2\right)\simeq
I\left(M_i^2/m_I^2\right)\sim0.1$ is assumed as can be seen from
Eq.~(\ref{eq:loop-int}). 
The result does not practically depend on $\Lambda_i$,
$M_2$ and $M_3$ 
since the dependence of $\Lambda_i$ appears in only
$\sigma_{\mathrm{eff}}^{\ell}v_{\mathrm{rel}}$ which has small
annihilation cross section. Thus we can see that the appropriate
contribution comes from the $t\overline{t}$-channel and $hh$-channel.
The red, green and blue points correspond to each
range of $\sin\alpha$ as written in Fig.~\ref{fig:plot}. 
In the case of $\left(y_S\right)_{11}=0.5$, the mass relation
$2M_1\approx m_H$ is required since the coupling $\left(y_S\right)_{11}$
is not so large and the annihilation cross section
$\sigma_{\mathrm{eff}}^{t}v_{\mathrm{rel}}$ has to be enhanced due to a
resonance. 
The resonance relation is not required for the right side one which
corresponds to $\left(y_S\right)_{11}=1.0$ if $\sin\alpha$ is large
since the $hh$-channel is effective in this case. 
There is no allowed parameter space in $M_1<m_h$ region because the main
channel is $\nu_{1\pm}\nu_{1\pm}\to b\overline{b}$ and the contribution
to the annihilation cross section is too small.

%%%%%%%%%%%%%%%%%%%%%%%%%%%%%
\section{Direct Detection}
\label{secdirect}

%%%%%%
The DM candidates $\nu_{1\pm}$ interact with quarks via Higgs exchange. 
Thus it is possible to explore the DM in direct detection experiments
like XENON100 \cite{xenon}. 
The Spin Independent (SI) elastic cross section $\sigma_{\mathrm{SI}}$
with nucleon $N$ is given by
\begin{equation}
\sigma_{\mathrm{SI}}^N\simeq\frac{\mu_{\mathrm{DM}}^2}{\pi}
\left(\frac{1}{m_h^2}-\frac{1}{m_H^2}\right)^2
\left(\frac{\left(y_S\right)_{11}m_N\sin\alpha\cos\alpha}
{\sqrt{2}v}\sum_{q}f_q^N\right)^2,
\label{eq:dd}
\end{equation}
where $\mu_{\mathrm{DM}}=\left(M_1^{-1}+m_N^{-1}\right)^{-1}$ is the
DM-nucleon reduced mass and the heavy Higgs contribution is neglected. 
The parameters $f_q^N$ which imply the contribution of each quark to
nucleon mass are calculated by the lattice
simulation~\cite{Corsetti:2000yq, Ohki:2008ff} as
\begin{eqnarray}
&&f_u^p=0.023,\quad
f_d^p=0.032,\quad
f_s^p=0.020,\\
&&f_u^n=0.017,\quad
f_d^n=0.041,\quad
f_s^n=0.020,
\end{eqnarray}
for the light quarks and $f_Q^N=2/27\left(1-\sum_{q\leq 3}f_q^N\right)$ for
the heavy quarks $Q$ where $q\leq3$ implies the summation of the light
quarks. The recent another calculation is performed in Ref.~\cite{Alarcon:2011zs}.

\begin{figure}[t]
\begin{center}
\includegraphics[scale=0.7]{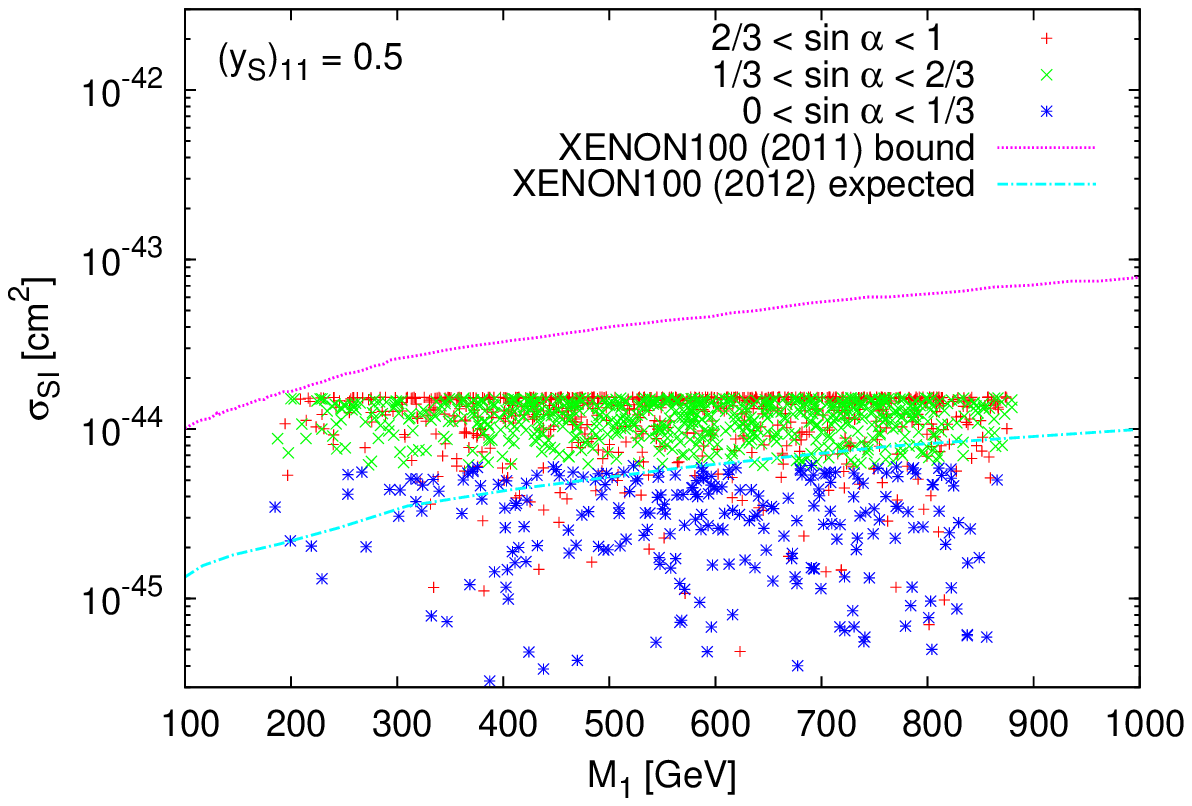}
\includegraphics[scale=0.7]{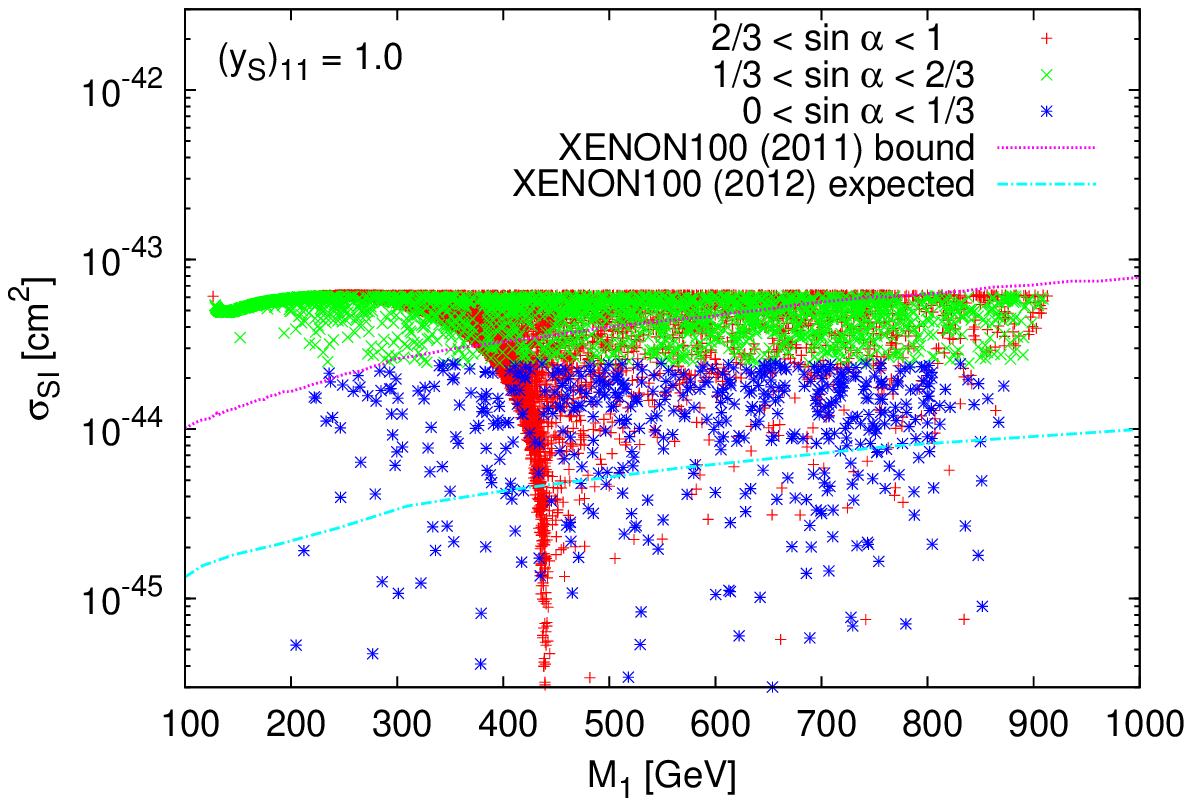}
\caption{
The comparison with XENON100 experiment. 
The left figure is for $\left(y_S\right)_{11}=0.5$ and the right one is
 for $\left(y_S\right)_{11}=1.0$. 
The parameter choice is same as Fig.~\ref{fig:plot}.}
\label{fig:dd}
\end{center}
\end{figure}

The comparison with XENON100 upper bound is shown in Fig.~\ref{fig:dd}
where the other parameters are fixed as same as Fig.~\ref{fig:plot} and
these correspond to red, green and blue points. The violet dotted line is XENON100 (2011) upper
bound and the light blue dashed line is XENON100 expected one in 2012. 
We can see that from the figure, the XENON100 (2011) limit excludes
$M_1\lesssim800~\mathrm{GeV}$ in the large $\sin\alpha$ region for
$\left(y_S\right)_{11}=1.0$. 
The almost excluded region of rather small $M_1$ implies that the parameter region of the
$hh$-channel is the most effective for the DM annihilation (Fig.~\ref{fig:plot}). 
The other certain region will be verified by the future XENON experiment. 
In the case of taking into account the lightest right-handed neutrino as
DM in the original radiative neutrino mass model~\cite{Ma:2006km}, the elastic
cross section with nuclei is not obtained at tree level because of the
leptophilic feature of the DM. However
interactions with quarks via Higgses are obtained in the radiative
inverse seesaw model and hence verification by direct detection of DM is possible. 

%%%%%%%%%%%%%%%%%%%%%%%%
\section{Conclusions}
We constructed a radiative inverse seesaw model which generates neutrino
masses and includes DMs simultaneously, and studied the mixing of
the lepton sector and the DM features.
The neutrino mass matrix was radiatively generated via the inverse seesaw
framework. 
Considering the latest data of non-zero $\theta_{13}$, we applied the
charged lepton mixing effect with almost Maximal Dirac CP phase
suggested by S. King. 
We can obtain the neutrino Dirac Yukawa matrix which produces the best
fit value of $\theta_{12}$ on the diagonal basis of the right-handed
neutrinos and additional fermions and non-zero $\theta_{13}$ comes from the
charged lepton mixing matrix. The size of the neutrino Yukawa couplings
is severely constrained by LFV at the same time. As a result, the
annihilation cross section which comes from the Yukawa interaction
becomes ineffective, however we found that new interactions via Higgs
bosons which are independent on LFV. Thus the DM can have the certain
annihilation cross section due to the interactions with Higgses. 
Verification of the model is also possible by direct detection of DM
through the interaction with Higgses. In particular, the region of the
large mixing $\sin\alpha$ will be testable by the exposure of the
XENON100 (2012) experiment. Therefore it is favored to be Higgs portal
DM in the radiative inverse seesaw model from the view point of avoiding
the LFV constraint and obtaining the proper detection rate by direct
search of DM.

%\newpage
%%%%%%%%%%%%%%%%%%%%%%%%%%%%%%%%%%%
%\vspace{0.5cm}
%\hspace{0.2cm} {\bf Acknowledgments}
\section*{Acknowledgments}
%\vspace{0.5cm}
H.O. thanks to Prof. Eung-Jin Chun, Dr. Priyotosh Bandyopadhyay, and
Dr. Jong-Chul Park, for fruitful discussion.
T. T. is supported by Young Researcher Overseas Visits Program for
Vitalizing Brain Circulation Japanese in JSPS. 
%%%%%%%%%%%%%%%%%%%%%%%%%%%%%%%%%%%
\newpage

\bibliographystyle{unsrt}

\begin{thebibliography}{99}
%%%%%%%%%%%%%%%%%%%%%%%%%%%%%%%%%%%%%%
%\cite{Mohapatra:1986aw}
\bibitem{Mohapatra:1986aw} 
  R.~N.~Mohapatra,
  %``Mechanism For Understanding Small Neutrino Mass In Superstring
	%Theories,''
  Phys.\ Rev.\ Lett.\  {\bf 56}, 561 (1986).
  %%CITATION = PRLTA,56,561;%%

%\cite{Mohapatra:1986bd}
\bibitem{Mohapatra:1986bd} 
  R.~N.~Mohapatra and J.~W.~F.~Valle,
  %``Neutrino Mass and Baryon Number Nonconservation in Superstring
	%Models,''
  Phys.\ Rev.\ D {\bf 34}, 1642 (1986).
  %%CITATION = PHRVA,D34,1642;%%
%\bibitem{inverse-origin}
%R. N. Mohapatra, Phys. Rev. Lett. {\bf 56}, 561 (1986);
%R. N. Mohapatra and J. W. F. Valle, Phys. Rev. D {\bf 34}, 1642 (1986).

%\cite{Abdallah:2011ew}
\bibitem{non-susy-inverse-pheno1} 
  W.~Abdallah, A.~Awad, S.~Khalil and H.~Okada,
  %``Muon Anomalous Magnetic Moment and mu -> e gamma in B-L Model with Inverse Seesaw,''
  arXiv:1105.1047 [hep-ph].
  %%CITATION = ARXIV:1105.1047;%%

%\cite{Khalil:2010iu}
\bibitem{non-susy-inverse-pheno2}
  S.~Khalil,
  %``TeV-scale gauged B-L symmetry with inverse seesaw mechanism,''
  Phys.\ Rev.\ D {\bf 82}, 077702 (2010)
  [arXiv:1004.0013 [hep-ph]].
  %%CITATION = ARXIV:1004.0013;%%

%\cite{Khalil:2011tb}
\bibitem{Khalil:2011tb} 
  S.~Khalil, H.~Okada and T.~Toma,
  %``Right-handed Sneutrino Dark Matter in Supersymmetric B-L Model,''
  JHEP {\bf 1107}, 026 (2011)
  [arXiv:1102.4249 [hep-ph]].
  %%CITATION = ARXIV:1102.4249;%%

%1-loop radiative seesaw
%ma model
%\cite{Ma:2006km}
\bibitem{Ma:2006km}
  E.~Ma,
  %``Verifiable radiative seesaw mechanism of neutrino mass and dark matter,''
  Phys.\ Rev.\  D {\bf 73}, 077301 (2006)
  [arXiv:hep-ph/0601225].
  %%CITATION = PHRVA,D73,077301;%%


%3-loop radiative seesaw
%\cite{Krauss:2002px}
\bibitem{Krauss:2002px}
  L.~M.~Krauss, S.~Nasri and M.~Trodden,
  %``A Model for neutrino masses and dark matter,''
  Phys.\ Rev.\  D {\bf 67}, 085002 (2003)
  [arXiv:hep-ph/0210389].
  %%CITATION = PHRVA,D67,085002;%%

%\cite{Aoki:2008av}
\bibitem{Aoki:2008av}
  M.~Aoki, S.~Kanemura and O.~Seto,
  %``Neutrino mass, Dark Matter and Baryon Asymmetry via TeV-Scale
	%Physics
  %without Fine-Tuning,''
  Phys.\ Rev.\ Lett.\  {\bf 102}, 051805 (2009)
  [arXiv:0807.0361].
  %%CITATION = PRLTA,102,051805;%%


%%%%% analysis of ma model
%%%%%   Since 2012 Jan.---  %%%%%%
%\cite{Schmidt:2012yg}
\bibitem{Schmidt:2012yg} 
  D.~Schmidt, T.~Schwetz and T.~Toma,
  %``Direct Detection of Leptophilic Dark Matter in a Model with
	%Radiative Neutrino Masses,''
  Phys.\ Rev.\ D {\bf 85}, 073009 (2012)
  [arXiv:1201.0906 [hep-ph]].
  %%CITATION = ARXIV:1201.0906;%%

%\cite{Bouchand:2012dx}
\bibitem{Bouchand:2012dx} 
  R.~Bouchand and A.~Merle,
  %``Running of Radiative Neutrino Masses: The Scotogenic Model,''
  arXiv:1205.0008 [hep-ph].
  %%CITATION = ARXIV:1205.0008;%%

%\cite{Ma:2012ez}
\bibitem{Ma:2012ez} 
  E.~Ma, A.~Natale and A.~Rashed,
  %``Scotogenic $A_4$ Neutrino Model for Nonzero $\theta_{13}$ and Large
	%$\delta_{CP}$,''
  arXiv:1206.1570 [hep-ph].
  %%CITATION = ARXIV:1206.1570;%%


%%%%% extended ma model
%%%%%   Since 2012 Jan.---  %%%%%%
%\cite{Aoki:2011he}
\bibitem{Aoki:2011he} 
  M.~Aoki, J.~Kubo, T.~Okawa and H.~Takano,
  %``Impact of Inert Higgsino Dark Matter,''
  Phys.\ Lett.\ B {\bf 707}, 107 (2012)
  [arXiv:1110.5403 [hep-ph]].
  %%CITATION = ARXIV:1110.5403;%%

%\cite{Ahn:2012cg}
\bibitem{Ahn:2012cg} 
  Y.~H.~Ahn and H.~Okada,
  %``Non-zero $\theta_{13}$ linking to Dark Matter from Non-Abelian
	%Discrete Flavor Model in Radiative Seesaw,''
  Phys.\ Rev.\ D {\bf 85}, 073010 (2012)
  [arXiv:1201.4436 [hep-ph]].
  %%CITATION = ARXIV:1201.4436;%%

%\cite{Ahn:2012cg, Farzan:2012sa}
\bibitem{Farzan:2012sa} 
  Y.~Farzan and E.~Ma,
  %``Scotogenic Dirac Neutrino Mass,''
  arXiv:1204.4890 [hep-ph].
  %%CITATION = ARXIV:1204.4890;%%

%\cite{Bonnet:2012kz}
\bibitem{Bonnet:2012kz} 
  F.~Bonnet, M.~Hirsch, T.~Ota and W.~Winter,
  %``Systematic study of the d=5 Weinberg operator at one-loop order,''
  arXiv:1204.5862 [hep-ph].
  %%CITATION = ARXIV:1204.5862;%%

%\cite{Kumericki:2012bf}
\bibitem{Kumericki:2012bf} 
  K.~Kumericki, I.~Picek and B.~Radovcic,
  %``Critique of Fermionic R\nuMDM and its Scalar Variants,''
  arXiv:1204.6597 [hep-ph].
  %%CITATION = ARXIV:1204.6597;%%

%\cite{Kumericki:2012bh}
\bibitem{Kumericki:2012bh} 
  K.~Kumericki, I.~Picek and B.~Radovcic,
  %``TeV-scale Seesaw with Quintuplet Fermions,''
  arXiv:1204.6599 [hep-ph].
  %%CITATION = ARXIV:1204.6599;%%

%\cite{Ma:2012if}
\bibitem{Ma:2012if} 
  E.~Ma,
  %``Radiative Scaling Neutrino Mass and Warm Dark Matter,''
  arXiv:1206.1812 [hep-ph].
  %%CITATION = ARXIV:1206.1812;%%


%\cite{Gil:2012ya}
\bibitem{Gil:2012ya} 
  G.~Gil, P.~Chankowski and M.~Krawczyk,
  %``Inert Dark Matter and Strong Electroweak Phase Transition,''
  arXiv:1207.0084 [hep-ph].
  %%CITATION = ARXIV:1207.0084;%%


%the other radiative neutrino mass models
%\cite{Aoki:2010ib}
\bibitem{Aoki:2010ib} 
  M.~Aoki, S.~Kanemura, T.~Shindou and K.~Yagyu,
  %``An R-parity conserving radiative neutrino mass model without
	%right-handed neutrinos,''
  JHEP {\bf 1007}, 084 (2010)
  [Erratum-ibid.\  {\bf 1011}, 049 (2010)]
  [arXiv:1005.5159 [hep-ph]].
  %%CITATION = ARXIV:1005.5159;%%

%\cite{Kanemura:2011vm}
\bibitem{Kanemura:2011vm} 
  S.~Kanemura, O.~Seto and T.~Shimomura,
  %``Masses of dark matter and neutrino from TeV scale spontaneous
	%$U(1)_{B-L}$ breaking,''
  Phys.\ Rev.\ D {\bf 84}, 016004 (2011)
  [arXiv:1101.5713 [hep-ph]].
  %%CITATION = ARXIV:1101.5713;%%

%\cite{Lindner:2011it}
\bibitem{Lindner:2011it} 
  M.~Lindner, D.~Schmidt and T.~Schwetz,
  %``Dark Matter and Neutrino Masses from Global $U(1)_{B-L}$ Symmetry
	%Breaking,''
  Phys.\ Lett.\ B {\bf 705}, 324 (2011)
  [arXiv:1105.4626 [hep-ph]].
  %%CITATION = ARXIV:1105.4626;%%

%\cite{Kanemura:2011mw}
\bibitem{Kanemura:2011mw} 
  S.~Kanemura, T.~Nabeshima and H.~Sugiyama,
  %``TeV-Scale Seesaw with Loop-Induced Dirac Mass Term and Dark Matter
	%from $U(1)_{B-L}$ Gauge Symmetry Breaking,''
  Phys.\ Rev.\ D {\bf 85}, 033004 (2012)
  [arXiv:1111.0599 [hep-ph]].
  %%CITATION = ARXIV:1111.0599;%%
%%%%%%%%%%%%%%%%%%




%\cite{Maki:1962mu}
\bibitem{mns}
  Z.~Maki, M.~Nakagawa and S.~Sakata,
  %``Remarks on the unified model of elementary particles,''
  Prog.\ Theor.\ Phys.\  {\bf 28}, 870 (1962).
  %%CITATION = PTPKA,28,870;%%


%\cite{Tortola:2012te}
\bibitem{Tortola:2012te} 
  M.~Tortola, J.~W.~F.~Valle and D.~Vanegas,
  %``Global status of neutrino oscillation parameters after recent
	%reactor measurements,''
  arXiv:1205.4018 [hep-ph].
  %%CITATION = ARXIV:1205.4018;%%


\bibitem{t2k} T2K Collaboration: K. Abe {\it et al.}, Phys. Rev. Lett.
{\bf 107}, 041801 (2011).
%\cite{Abe:2011fz}
\bibitem{Abe:2011fz} 
  Y.~Abe {\it et al.}  [DOUBLE-CHOOZ Collaboration],
  %``Indication for the disappearance of reactor electron antineutrinos
	%in the Double Chooz experiment,''
  Phys.\ Rev.\ Lett.\  {\bf 108}, 131801 (2012)
  [arXiv:1112.6353 [hep-ex]].
  %%CITATION = ARXIV:1112.6353;%%

\bibitem{daya} Daya Bay Collaboration: F. P. An {\it et al.},
arXiv:1203.1669 [hep-ex].

\bibitem{reno} RENO Collaboration: J. K. Ahn {\it et al.},
arXiv:1204.0626 [hep-ex].


%\cite{Antusch:2005kw}
\bibitem{king} 
  S.~Antusch and S.~F.~King,
  %``Charged lepton corrections to neutrino mixing angles and CP phases revisited,''
  Phys.\ Lett.\ B {\bf 631}, 42 (2005)
  [hep-ph/0508044];
  %%CITATION = HEP-PH/0508044;%%
%\cite{King:2012vj}
\bibitem{King:2012vj} 
  S.~F.~King,
  %``Tri-bimaximal-Cabibbo Mixing,''
  arXiv:1205.0506 [hep-ph].
  %%CITATION = ARXIV:1205.0506;%%
  


\bibitem{wmap}
  E.~Komatsu {\it et al.}  [WMAP Collaboration],
  %``Seven-Year Wilkinson Microwave Anisotropy Probe (WMAP) Observations:
  %Cosmological Interpretation,''
  Astrophys.\ J.\ Suppl.\  {\bf 192}, 18 (2011)
  [arXiv:1001.4538 [astro-ph.CO]].
  %%CITATION = APJSA,192,18;%%

%\cite{Aprile:2011hi}
\bibitem{xenon} 
  E.~Aprile {\it et al.}  [XENON100 Collaboration],
  %``Dark Matter Results from 100 Live Days of XENON100 Data,''
  Phys.\ Rev.\ Lett.\  {\bf 107}, 131302 (2011)
  [arXiv:1104.2549 [astro-ph.CO]].
  %%CITATION = ARXIV:1104.2549;%%


%\bibitem{prepare}
%Higgs Signature in Inverse Seesaw Model at the LHC
% Priyotosh Bandyopadhyay, Eung Jin Chun, Hiroshi Okada and Jong-chul Park, in preparing.

%\cite{Ma:2009gu}
\bibitem{Ma:2009gu} 
  E.~Ma,
  %``Radiative inverse seesaw mechanism for nonzero neutrino mass,''
  Phys.\ Rev.\ D {\bf 80}, 013013 (2009)
  [arXiv:0904.4450 [hep-ph]].
  %%CITATION = ARXIV:0904.4450;%%

%\cite{Carena:2004xs}
\bibitem{Carena:2004xs} 
  M.~S.~Carena, A.~Daleo, B.~A.~Dobrescu and T.~M.~P.~Tait,
  %``$Z^\prime$ gauge bosons at the Tevatron,''
  Phys.\ Rev.\ D {\bf 70}, 093009 (2004)
  [hep-ph/0408098].
  %%CITATION = HEP-PH/0408098;%%


%neutrinoless double beta decay
%\cite{Rodejohann:2011mu}
\bibitem{Rodejohann:2011mu} 
  W.~Rodejohann,
  %``Neutrino-less Double Beta Decay and Particle Physics,''
  Int.\ J.\ Mod.\ Phys.\ E {\bf 20}, 1833 (2011)
  [arXiv:1106.1334 [hep-ph]].
  %%CITATION = ARXIV:1106.1334;%%


%\cite{Adam:2011ch}
\bibitem{Adam:2011ch} 
  J.~Adam {\it et al.}  [MEG Collaboration],
  %``New limit on the lepton-flavour violating decay $\mu^{+} \to e^{+}
	%\gamma$,''
  Phys.\ Rev.\ Lett.\  {\bf 107}, 171801 (2011)
  [arXiv:1107.5547 [hep-ex]].
  %%CITATION = ARXIV:1107.5547;%%

%\cite{Nakamura:2010zzi}
\bibitem{Nakamura:2010zzi} 
  K.~Nakamura {\it et al.}  [Particle Data Group Collaboration],
  %``Review of particle physics,''
  J.\ Phys.\ G G {\bf 37}, 075021 (2010).
  %%CITATION = JPHGB,G37,075021;%%

%%%%
%\cite{Kubo:2006yx}
\bibitem{Kubo:2006yx} 
  J.~Kubo, E.~Ma and D.~Suematsu,
  %``Cold Dark Matter, Radiative Neutrino Mass, mu ---> e gamma, and
	%Neutrinoless Double Beta Decay,''
  Phys.\ Lett.\ B {\bf 642}, 18 (2006)
  [hep-ph/0604114].
  %%CITATION = HEP-PH/0604114;%%

%\cite{Kubo:2006rm}
\bibitem{Kubo:2006rm} 
  J.~Kubo and D.~Suematsu,
  %``Neutrino masses and CDM in a non-supersymmetric model,''
  Phys.\ Lett.\ B {\bf 643}, 336 (2006)
  [hep-ph/0610006].
  %%CITATION = HEP-PH/0610006;%%

%\cite{Suematsu:2009ww}
\bibitem{Suematsu:2009ww} 
  D.~Suematsu, T.~Toma and T.~Yoshida,
  %``Reconciliation of CDM abundance and mu ---> e gamma in a radiative
	%seesaw model,''
  Phys.\ Rev.\ D {\bf 79}, 093004 (2009)
  [arXiv:0903.0287 [hep-ph]].
  %%CITATION = ARXIV:0903.0287;%%

%\cite{Suematsu:2010nd}
\bibitem{Suematsu:2010nd} 
  D.~Suematsu and T.~Toma,
  %``Dark matter in the supersymmetric radiative seesaw model with an
	%anomalous U(1) symmetry,''
  Nucl.\ Phys.\ B {\bf 847}, 567 (2011)
  [arXiv:1011.2839 [hep-ph]].
  %%CITATION = ARXIV:1011.2839;%%



%\cite{Corsetti:2000yq}
\bibitem{Corsetti:2000yq} 
  A.~Corsetti and P.~Nath,
  %``Gaugino mass nonuniversality and dark matter in SUGRA, strings and
	%D-brane models,''
  Phys.\ Rev.\ D {\bf 64}, 125010 (2001)
  [hep-ph/0003186].
  %%CITATION = HEP-PH/0003186;%%

%\cite{Ohki:2008ff}
\bibitem{Ohki:2008ff} 
  H.~Ohki, H.~Fukaya, S.~Hashimoto, T.~Kaneko, H.~Matsufuru, J.~Noaki,
	T.~Onogi and E.~Shintani {\it et al.},
  %``Nucleon sigma term and strange quark content from lattice QCD with
	%exact chiral symmetry,''
  Phys.\ Rev.\ D {\bf 78}, 054502 (2008)
  [arXiv:0806.4744 [hep-lat]].
  %%CITATION = ARXIV:0806.4744;%%

%\cite{Alarcon:2011zs}
\bibitem{Alarcon:2011zs} 
  J.~M.~Alarcon, J.~Martin Camalich and J.~A.~Oller,
  %``The chiral representation of the $\pi N$ scattering amplitude and the pion-nucleon sigma term,''
  Phys.\ Rev.\ D {\bf 85}, 051503 (2012)
  [arXiv:1110.3797 [hep-ph]].
  %%CITATION = ARXIV:1110.3797;%%

\if0
%%%%% Added in updating  %%%%
%\cite{Chiang:2012qa}
\bibitem{Chiang:2012qa} 
  C.~-W.~Chiang, T.~Nomura and J.~Tandean,
  %``Dark Matter and Higgs Boson in a Model with Discrete Gauge Symmetry,''
  arXiv:1205.6416 [hep-ph].
  %%CITATION = ARXIV:1205.6416;%%
\fi

\end{thebibliography}

\end{document}